\documentclass{emulateapj}
\usepackage{graphicx,multirow,txfonts,hyperref,url,color}
\hypersetup{colorlinks=true, urlcolor=blue, citecolor=blue}

\def\be{\begin{equation}}
\def\ee{\end{equation}}
\newcommand{\kev}{{\rm \,keV}}

\newcommand{\ergs}{{\rm\,erg\,s^{-1}}}
\newcommand{\msun}{{\rm M}_{\sun}}

\newcommand{\tc}{T_{\rm C}}
\catcode`\@=11 % This allows us to modify PLAIN macros.
\def\@versim#1#2{\vcenter{\offinterlineskip
        \ialign{$\m@th#1\hfil##\hfil$\crcr#2\crcr\sim\crcr } }}

\begin{document}
%\date{}

%\slugcomment{Submitted to The Astrophysical Journal}
%\title{Compton Temperature of low-luminosity Active Galactic Nuclei}
%\shorttitle{Compton Temperature of LLAGNs}
\title{Radiative heating in the kinetic mode of AGN feedback}
\shorttitle{Radiative heating in kinetic mode of AGN feedback}

\author{Fu-Guo Xie$^1$, Feng Yuan$^1$, and Luis C. Ho$^{2,3}$}
\affil{$^1$ Key Laboratory for Research in Galaxies and Cosmology, Shanghai Astronomical Observatory, \\
Chinese Academy of Sciences, 80 Nandan Road, Shanghai 200030, China; \href{mailto:fgxie@shao.ac.cn}{fgxie},\href{mailto:fyuan@shao.ac.cn}{fyuan@shao.ac.cn}\\
$^2$ Kavli Institute for Astronomy and Astrophysics, Peking University, Beijing 100871, China; \href{mailto:lho.pku@gmail.com}{lho.pku@gmail.com}\\
$^3$ Department of Astronomy, School of Physics, Peking University, Beijing 100871, China}
\shortauthors{Xie, Yuan \& Ho}

\begin{abstract}
AGN feedback is now widely believed to play a crucial role in the co-evolution between the central black hole and its host galaxy. Two feedback modes have been identified, namely the radiative and kinetic modes, which correspond to the luminous AGNs and low-luminosity AGNs (LLAGNs), respectively. In this paper, we investigate the radiative heating in the kinetic mode. This process is potentially important because: 1) the radiation power of LLAGNs is higher than the jet power over a wide parameter range; 2) the spectral energy distribution of LLAGNs is such that the radiative heating is more effective compared to that of luminous AGNs with the same luminosity; and 3) most of the time in the lifecycle of an AGN is spent in the LLAGNs phase. In this paper, adopting the characteristic broad-band spectral energy distributions of LLAGNs, we calculate the value of ``Compton temperature'' ($\tc$), which determines the radiative heating by Compton scattering. We find that $\tc\sim (5-15)\times 10^7$ K, depending on the spectrum of individual LLAGN and at which distance from the black hole we evaluate the heating. We also compare this heating process with other radiative heating and cooling processes such as photoionization/recombination. Our result can be used for an accurate calculation of the radiative heating in the study of AGN feedback.
\end{abstract}

\keywords{galaxies: Seyfert --- galaxies: active --- accretion, accretion disks}

\section{Introduction}\label{sec:intro}

There are considerable observational evidences for the co-evolution of the supermassive black hole and its host galaxy, and the co-evolution is now widely believed to be due to active galactic nuclei (AGNs) feedback (e.g., \citealt{Magorrian98, Gebhardt00, Fabian12, KH13, HB14}). Accretion onto the supermasive black hole in the galactic center will produce both radiation and outflows. These outputs will interact with the interstellar medium (ISM) in the host galaxy, near or far from the black hole, by transferring their momentum and energy to the ISM. The gas will then be heated up or pushed away from the black hole. The changes in the temperature and density of the gas, on one hand, will obviously affect the star formation and galaxy evolution. On the other hand, they will also affect the fueling of the black hole by changing the accretion rate, thus the radiation and matter output of accretion, and the growth of the black hole mass.

While this field is still relatively young and there are many unsolved problems, some consensus has been reached. Two feedback modes have been identified, which correspond to two accretion modes \citep{Fabian12, KH13, HB14}. One is called the radiative or quasar mode. This mode operates when the black hole accretes at a significant fraction of the Eddington rate. In this case, the accretion flow is in the standard thin disk regime \citep{SS73} and the corresponding AGNs are very luminous. The other mode is called the kinetic mode or radio mode or maintenance mode, when the black hole accretes at a low accretion rate. In this case, the accretion flow is described by a hot accretion flow \citep{NY94, Yuan14}. The corresponding AGNs are called low-luminosity AGNs (LLAGNs). By analogy with the soft and hard states of black hole X-ray binaries (BHBs) (see \citealt{Belloni10} for the classification of states in BHBs), the boundary between the two modes is $L_{\rm bol}\sim (1-2)\% L_{\rm Edd}$, where $L_{\rm bol}$ is the bolometric luminosity and $L_{\rm Edd}\approx 1.3\times10^{46} (M_{\rm BH}/10^8 \msun)\ \ergs$ is the Eddington luminosity.

The output of black hole accretion generally includes three components, i.e., radiation, jet, and wind. The difference between the latter two is that jet has relativistic speed and is well-collimated, while wind is sub-relativistic and has a much larger solid angle.  In the radiative mode of feedback, i.e., the standard thin disk case, there is no jet. In addition to the radiation which is obviously strong, we have also observational evidence of wind, e.g., in the case of broad-absorption-line (BAL) quasars (see \citealt{Crenshaw03} for review of observations). These winds may be driven by the radiation line force (e.g., \citealt{Murray95, Proga00}).

In the kinetic mode of feedback, all three kinds of output exist. Among them, jet is perhaps most widely considered in the study, mainly because observationally jets are most evident (e.g., \citealt{Ho08}). However, it is still under active debate whether jet at large scales should be described by hydrodynamic (e.g., \citealt{Guo16}), magnetohydrodynamic (e.g., \citealt{Gan17} and references therein), or cosmic-ray dominated one \citep{Guo11}, and more importantly, how efficient the jet can deposit its energy into the ISM or intergalactic medium because of its very small solid angle and rather high velocity (e.g., \citealt{Vernaleo06}).

In the study of black hole accretion, wind from hot accretion flow is one of the most important progresses in recent years. Here we only briefly summarize the main development and readers are referred to the recent review by \citet{Yuan16} for more details. In the pioneer work of \citet{NY94}, it has been speculated that strong outflow should be easily formed because of the positive Bernoulli parameter of the hot accretion flow. \citet{BB99} proposed an analytical model by emphasizing the wind. Since wind in the accretion flow is intrinsically a multi-dimensional physical process, the proper study of wind can only be achieved through numerical simulations. \citet{Stone99} has performed the first global numerical simulation of black hole accretion and found that the mass accretion rate decreases with decreasing radius. By analyzing the numerical simulation data, including analyzing the convective instability of an MHD accretion flow, \citet{Yuan12} convincingly showed that such a decrease in accretion rate must be caused by strong wind rather than convection (see also \citealt{Narayan12, Li13}). By using a ``virtual particle trajectory'' approach, \citet{Yuan15} have carefully studied the properties of wind. They find that generally the mass flux of wind is much larger than the accretion rate, and fluxes of energy and momentum are also much larger than that of jet in the case of accretion onto a Schwarzschild black hole\footnote{In this case, of course there is no \citet[][BZ]{Blandford77} jet. But simulations have shown that there exists a ``disk-jet'', which is powered by the rotating accretion flow. The differences between the disk-jet and BZ-jet are discussed in \citet{Yuan14} and \citet{Yuan15}.}. Because the wind gas is fully ionized, it is very difficult to detect wind from observing absorption lines. Still, more and more observations confirm the existence of wind in both LLAGNs \citep{Cheung16, Crenshaw12, Tombesi14} and the hard state of black hole X-ray binaries \citep{Homan16}, where a hot accretion flow is believed to operate. In many works, winds have been
included although without explicitly emphasizing the mode of
feedback (e.g., \citealt{Ostriker10, Ciotti10, Eisenreich17}). Achieving a sufficiently rapid reddening of moderately massive galaxies without expelling too many baryons has been challenging for simulations of galaxy formation. Most recently, by invoking the kinetic feedback effect from winds in the regime of low accretion rates (i.e., the kinetic feedback mode), \citet{Weinberger17} has successfully solved this problem.

In this work, we focus on another mechanism in the kinetic mode, i.e. the radiative heating. This feedback mechanism is ignored sometimes in previous works (but see e.g., \citealt{Ciotti01, Ostriker10, Choi12, Gan14, Eisenreich17}), perhaps because it is thought that the radiation of hot accretion flow is too weak. However, based on the following reasons, it is necessary to study its potential role in feedback. First, the luminosity of a hot accretion flow covers a very wide range depending on the accretion rate, and can be moderately high. Taking the black hole X-ray binary as an example, the hard and soft states are described by the hot accretion flow and the thin disk, respectively \citep{MR06, RM06, Done07}. The highest luminosity of the hard state can be $L_{\rm bol}\sim (2-10)\%\ L_{\rm Edd}$ (e.g., \citealt{MR06, RM06, Done07}). Theoretically, the high luminosity of a hot accretion flow is because: 1) the radiative efficiency is a function of accretion rate, it increases with the increasing accretion rate \citep{Xie12}; 2) the highest accretion rate of a hot accretion flow can be $\ga 10^{-2}\dot{M}_{\rm Edd}$ \citep{Yuan14}. Compared with the power of a jet, the power of radiation will be larger when the X-ray luminosity $L_{\rm X} \ga 4\times 10^{-5}L_{\rm Edd}$ (or roughly the bolometric luminosity $L_{\rm bol}\ga 6\times 10^{-4}L_{\rm Edd}$) \citep{Fender03}. Second, the spectrum emitted by a hot accretion flow is different from that by a thin disk. As summarized by \citet[][see also \citealt{Ho08}]{Ho99}, the main difference of the spectrum of LLAGNs from luminous AGNs is the lack of the big blue bump. This means that for a given luminosity, there will be more hard photons. This results in a more effective radiative heating, as we will see later from Equations (8) or (9). Finally, galactic nuclei spend most of their time in the LLAGN phase rather than in the active phase (e.g., \citealt{HR93, Kauffmann00, MW01}). Thus the cumulative effect of radiative heating in the kinetic mode may be significant.

Radiative heating mainly include two processes. One is heating by Compton scattering; the other is by photoionization. We will see that the former is typically determined by the spectrum of the LLAGNs; the dependence on the properties of the ISM is very weak. On the other hand, the latter is a strong function of the ionization parameter $\xi$, which is sensitive to the local properties of the ISM. Moreover, the calculation of the latter is relatively straightforward. Because of these reasons, in this paper we focus on the Compton heating. In principle, Compton scattering can be either a heating or a cooling process, depending on the energy contrast between the photons and electron; but in practice, we will see that it is usually a heating process. We will evaluate the Compton heating rate by using ``Compton temperature'' ($\tc$), following the approach of \citet{Sazonov04}.  Physically, Compton temperature means the gas temperature at which net energy exchange by Compton scattering between photons and electrons vanishes, and it is determined by the energy-weighted average energy of the emitted photons from LLAGNs, cf. Equation (\ref{eq:tcdefine}) below. \citet{Sazonov04} calculated the Compton temperature of typical luminous AGNs and found $\tc\approx 2\times 10^7$ K. The main aim of the present work is to calculate  the value of $\tc$ of LLAGNs.  For this aim, in \S2, we combine the data from literature to obtain the broad-band spectral energy distribution of  LLAGNs. Special attention will be paid to the hard X-ray spectrum since this is the most important part in the spectrum for heating. We then in \S3 calculate the corresponding Compton temperature. In \S4 we compare the Compton heating with the other heating and cooling processes, such as photoionization heating, recombination and line cooling, and bremsstrahlung cooling, to see the relative importance of Compton heating. The final section is devoted to discussions and a short summary.

%{\bf Among numerous numerical simulations on AGN feedback (e.g., \citealt{Ciotti01, Ciotti07, Ciotti12, Ciotti10, Ostriker10, Choi12, Gan14, Eisenreich17}),} winds have been included in many of them although without explicitly emphasizing its mode of feedback

\begin{table*}
\begin{center}
\vspace{0.1 cm}
\centerline{Table 1 -- LLAGNs with observational constrains on $E_{\rm c}$}\label{tab1}
\vspace{0.2 cm}	
 \begin{tabular}{l l c c c c c c l}	
  \hline\hline
Name     &      Class   &       distance   & $M_{\rm BH}$ & Notes on & $L_{\rm X}/L_{\rm Edd}$ &  $\Gamma$  & $E_c$ & References on X-ray \\	
             &           &     (Mpc)       & ($\msun$) & $M_{\rm BH}$ &  &  & ($\kev$) & properties ($L_{\rm X}/L_{\rm Edd}, \Gamma, E_c$) \\
\hline
M~87 & LINER & 16.7 & $3.5\ \times 10^9$ & Dyn (W13) & $8.1\ \times 10^{-8}$ & $2.17\pm 0.01$	&  $>1000$ & \citealt{WY02}\\
NGC~4151    &   Sy 1  &   19   & $3.8\ \times 10^7$  & Dyn (O14) &  $2.\ \times 10^{-3}$   &   $1.77^{+0.06}_{-0.05}$ &  $307^{+245}_{-94}$ & \citealt{Molina09} \\
 &     &   &  &  & ... & $1.85\pm0.09$  & $450^{+900}_{-200}$ & \citealt{Beckmann05}\\
 &     &   &  &  &  $1.4\times10^{-3}$   & $1.81^{+0.05}_{-0.03}$  & $>1025$ & \citealt{Lubinski10}\\
 &     &   &  &  & $7.\times10^{-4}$   & $1.81\pm0.01$  & $>1325$ & \citealt{Lubinski10}\\
MCG-06-30-15 & Sy 1.2& 37.4 & $1.2\ \times 10^8$ & $M_{\rm BH}$-$\sigma$ & $3.2\ \times 10^{-4}$ & $2.18^{+0.10}_{-0.11}$ & $>76$ & \citealt{Molina09} \\
NGC~4593 & Sy 1 & 43.5 & $1.0\ \times 10^7$ & RM (D06) & $4.7\ \times 10^{-3}$ & $1.92\pm 0.01$ & $>222$ & \citealt{Molina09}\\
& & & & & $3.1\times10^{-3}$ & $1.84\pm 0.01$ & $>640$ & \citealt{Ursini16}\\
NGC~6814 & Sy 1.5 & 22.6 & $2.4\ \times 10^7$ & $M_{\rm BH}$-$\sigma$ & $3.2\ \times 10^{-5}$ & $1.80\pm 0.09$ & $116^{+203}_{-53}$ & \citealt{Molina09}\\
MCG-02-58-022 & Sy 1.5 & 201 & $3.9\ \times 10^8$ & $M_{\rm BH}$-$\sigma$ & $2.8\ \times 10^{-3}$ & $1.95^{+0.03}_{-0.04}$ & $>510$ & \citealt{Bianchi04, Malizia14}\\
NGC~7213 & LINER & 21.2 & $1\ \times 10^8$ & $M_{\rm BH}$-$\sigma$ & $1.5\ \times 10^{-4}$ & 1.85 & $>350$ & \citealt{Emm12, Lobban10}\\
&   &  &  &   & $1.0\ \times 10^{-4}$ & $1.84 \pm 0.03$ & $>140$ & \citealt{Ursini15}\\
NGC~5506 & Sy 1.9 & 29.1 & $2\ \times 10^8$ & $M_{\rm BH}$-$\sigma$ & $3.4\ \times 10^{-4} $ & $1.9$ & $720^{+130}_{-190}$ & \citealt{Matt15}\\
Cen~A & radio gal. & 3.8 & $5\ \times 10^7$ & Dyn (N07) & $1.2\times10^{-4}$ & $1.67$ & $>700$ & \citealt{Burke14}\\
& & & & &  $2.5\times10^{-4}$ & $1.815\pm0.005$ & $>1000$ & \citealt{Furst16}\\
& & & & &  $3.9\times 10^{-4}$ & $1.73\pm0.02$ & $434_{-73}^{+106}$ & \citealt{Beckmann11}\\
\hline																						
\end{tabular}
\end{center}
\small									
Notes. \\
1) The distance of M87 is derived from Tully-Fisher relationship \citep{Neill14}, that of NGC~4151 is from the parallax method \citep{Honig14}, and that of Cen A is the best-estimate based on various redshift-independent distance measures \citep{Harris10}. The distance of rest sources, on the other hand, is derived from redshift, assuming a flat cosmology. \\
2) Various methods are adopted to measure/estimate the black hole mass, i.e. stellar or gas dynamics (marked by ``Dyn''), the reverberation-mapping (marked by ``RM'') or $M_{\rm BH}$-$\sigma$ relationship (marked by ``$M_{\rm BH}$-$\sigma$'', with formulae taken from \citet{KH13} and the velocity dispersion from website http://leda.univ-lyon1.fr.).\\
References on $M_{\rm BH}$ measurements. W13: \citealt{Walsh13}; O14: \citealt{Onken14}; D06: \citealt{Denney06}; N07: \citealt{Neumayer07}.
\vspace{0.3cm}
 \end{table*}

\section{Broad-band spectrum of LLAGNs}\label{sec:sed}

In the kinetic feedback mode, i.e., when the luminosity of the AGNs is $\la (1-2)\%L_{\rm Edd}$, the broad-band spectrum of the AGNs has qualitative differences from that of a luminous AGN, with the most significant one being the absence of the `big-blue-bump'' that is present in the spectrum of luminous AGNs \citep{Ho99, Ho08}. Theoretically, this is because in luminous AGNs a standard thin disk extends to the innermost stable circular orbit, while in LLAGNs it is truncated at a transition radius and replaced by a hot accretion flow within this radius \citep{Yuan14}.  Another notable feature is that since the luminosity of LLAGNs covers a large range, the spectrum at different luminosity regime is also different.  In this section, we investigate the broad-band spectrum of LLAGNs at various luminosity regimes.
%In this section, we combine data from literature to obtain a broad-band spectrum at various luminosity regimes.
For the calculation of Compton heating, the spectrum in the hard X-ray band is crucial, so we first discuss the hard X-ray spectrum of LLAGNs.

\subsection{Hard X-ray spectrum of LLAGNs: Photon index and cutoff energy}\label{sec:ecutoff}

The hard X-ray and Gamma-ray emission of LLAGNs are of crucial importance to determine their Compton temperatures. In practice, the X-ray spectrum can be well described by a power-law with an exponential cutoff,
\begin{equation}
F_E \propto E^{1-\Gamma} \exp (-E/E_c),\label{eq:xsed}
\end{equation}
where $\Gamma$ is the photon index of the hard X-ray spectrum, $E\equiv h\nu$ is the photon energy and $E_c$ is the exponential cutoff energy (or the $e$-folding energy). Additionally there may also exist a reflection component in energy band $10 - 50\ \kev$. Observationally the X-ray photon index $\Gamma$ is now measured fairly well. The value of $\Gamma$ in LLAGNs generally locates in the range $\Gamma\approx 1.5 - 1.9$, and it anti-correlates with the X-ray luminosity $L_{\rm X}/L_{\rm Edd}$ (e.g., \citealt{Yang15, Emm12, Connolly16}).

The cutoff energy $E_c$, on the other hand, remains poorly constrained. Only dozens of AGNs have such measurements, mainly thanks to recent advances in hard X-ray ($E\ga 30-50 \kev$) telescopes and instruments, i.e. {\it CGRO}/OSSE (e.g. \citealt{Maisack93, Zdziarski95, Zdziarski00, Gondek96}), {\it BeppoSAX}/PDS (e.g. \citealt{Perola02, DR07, Dadina08}), {\it Integral}/IBIS/ISGR (e.g. \citealt{Beckmann05, Beckmann09, Panessa08, Malizia08, Molina09, Lubinski10, Lubinski16, Beckmann11, Molina13, Malizia12, Malizia14}), {\it Swift}/BAT (e.g. \citealt{Winter09, Burlon11, Molina13}), and {\it NuSTAR} (e.g. \citealt{Brenneman14, Marinucci14, Matt15, Ursini15, Ursini16, Furst16}). Statistically,  from {\it BeppoSAX} observations \citet{Dadina08} found that the nearby ($z<0.1$) Seyfert galaxies (105 objects in total), on average, have photon index $\Gamma\sim 1.8$ and cutoff energy $E_c\sim290\ \kev$. Similar result has also been obtained by \citet{Beckmann09} based on {\it Integral} observations. However, we note that most of the sources currently explored are moderately bright (in Eddington unit) and belong to the luminous AGN category. Besides, a reliable measurement of $E_c$ requires broadband spectral studies, which implies that ideally both low- and high-energy X-ray spectra have to be observed and modelled simultaneously, employing spectra with high statistical quality such as those acquired. This difficulty further limits the number of sources with reliable measurements of $E_c$.

Compared to luminous AGNs, the $e$-folding cutoff energy $E_c$ of LLAGNs is much more difficult to constrain, because of their systematically lower X-ray flux. Considering the uncertainties, we gather from literature (mainly select from \citealt{Molina09, Malizia12, Malizia14}, see Table\ \ref{tab1} for references of individual sources) LLAGNs which satisfy $L_{\rm X} \lesssim 4\times 10^{-3}\ L_{\rm Edd}$. There are nine sources in total, as summarised in Table\ \ref{tab1}. We include in this table the source name, AGN classification, black hole mass, distance, X-ray luminosity, photon index and cutoff energy. As noted in the table, the black hole mass $M_{\rm BH}$ are calculated through various methods. Besides, as noted in Table\ \ref{tab1}, there are redshift-independent distance measurements on the distance for several nearby sources, while for the rest the distance is calculated from redshift in a flat cosmology with $H_0 = 70\ {\rm km ~s^{-1} ~Mpc^{-1}}, \Omega_M = 0.27, \Omega_\Lambda = 0.73$. Due to the lack of sensitive instruments in the $200-800\ \kev$ energy band, most of the LLAGNs only have a lower limit constraint on $E_c$.

One of the best $E_c$ measurement comes from LLAGN NGC~7213 \citep{Emm12}, which is classified as a low-ionization nuclear emission-line region (LINER). An anti-correlation between $\Gamma$ and $L_{\rm X}/L_{\rm Edd}$ (the so-called ``harder when brighter'' behaviour) is observed in this source \citep{Emm12}. The cutoff energy is constrained to be $E_c > 350\ \kev$ by {\it Suzaku} and {\it Swift}/BAT \citep{Lobban10}, or $E_c > 140\ \kev$ by {\it NuSTAR} \citep{Ursini15}.

Another example is NGC~5506, which is classified as either a Seyfert 1.9 or a narrow-line Seyfert 1 galaxy. The X-ray photon index, with a typical value $\Gamma\approx 1.9$ \citep{Bianchi04, Matt15}, also anti-correlates with the X-ray luminosity \citep{Soldi14}. From simultaneous XMM-{\it Newton}/{\it BeppoSAX} observations, \citet{Bianchi04} found that the cutoff energy is $E_c = 140^{+40}_{-30}\ \kev$. However, as pointed out recently by \citet{Matt15}, this value suffers large systematic uncertainties due to ambiguities during the spectral modelling. Indeed, the {\it NuSTAR} observation on this  source found that $E_c = 720^{+130}_{-190}\ \kev$ \citep{Matt15}. Even allowing for systematic uncertainties, they confirm the $3\sigma$ lower limit of the cutoff to be $E_c > 350\ \kev$.

According to Table\ \ref{tab1}, we may set $E_c = 400 - 800\ {\rm keV}$ for LLAGNs. Besides, the value of $E_c$ might anti-correlate with the X-ray luminosity $L_{\rm X}/L_{\rm Edd}$, as indicated by observations of one LLAGN (NGC 4593,  \citealt{Ursini16}) and BHBs in their hard state (among others see e.g., GX 339-4, cf. Fig. 7 in \citealt{Miyakawa08}),  and expected by theory of hot accretion flows.

\begin{figure*}
\centering
\includegraphics[width=14. cm]{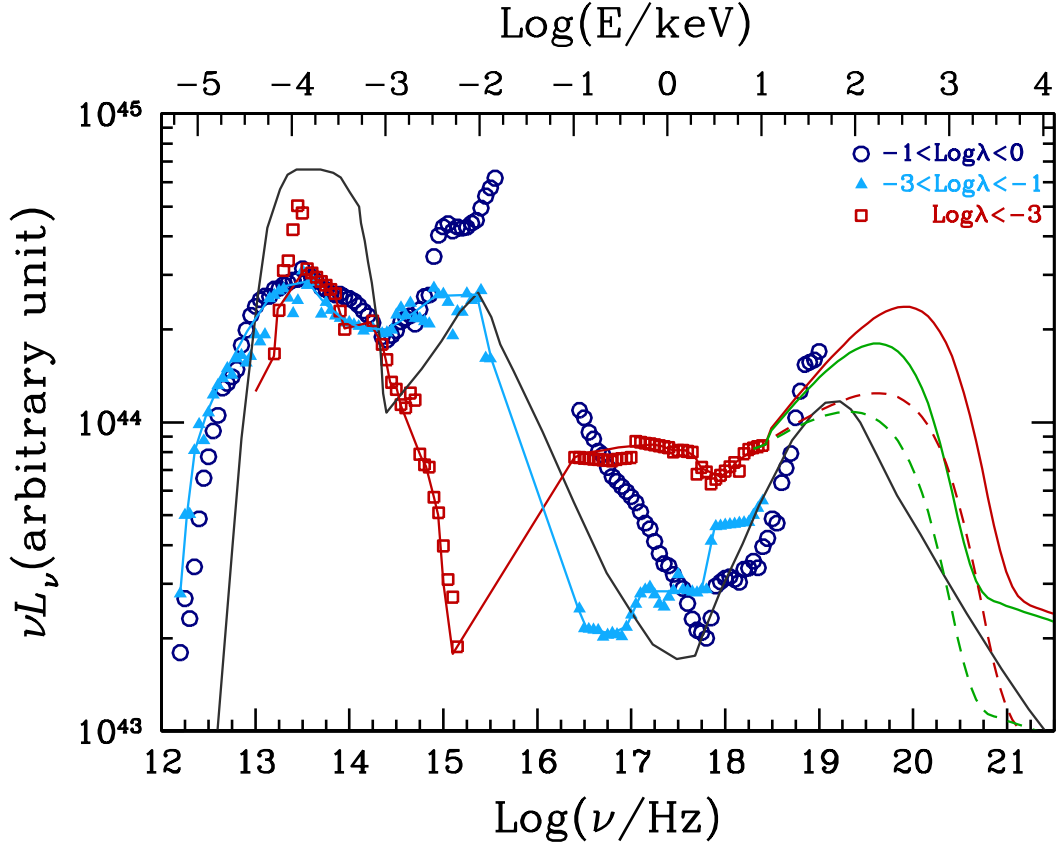}
\vspace{-0.1 cm}
\caption{Composite SED of AGNs. The SEDs are normalized at 1 $\mu$m (equivalently $3\times10^{14}$ Hz). The black solid curve is from \citet{Sazonov04} (averaged over Type 1 and Type 2 AGNs), and the rest observational data are taken from \citet{Ho08}, with Eddington luminosity ratio $\lambda$ labeled in the figure. In order to investigate the Compton heating effect, we complete the high-energy band SED ($E>10\ \kev$) of low-luminosity AGNs with several possibilities, i.e. the solid curves have relatively hard spectra, with $\Gamma = 1.60$, while the dashed curves have relatively soft spectrum, with $\Gamma = 1.80$. Different colors indicate different $e$-folding cutoff energies, i.e. dark green and dark red are respectively for $E_c = 400\ \kev$ and $E_c = 800\ \kev$.}\label{fig:LLAGN_SED}
\end{figure*}

\subsection{Composite broad-band SED of LLAGNs}

Obtaining the broad-band spectrum of LLAGNs is challenging. Various sample selection and normalization methods have been developed (e.g. \citealt{Ho99,Ho08,Malizia03,Winter09,Eracleous10}). As shown in Fig.\ \ref{fig:LLAGN_SED}, we here adopt the composite SED of LLAGNs from \citet{Ho08}, which has a relatively broad coverage in photon energy, i.e. from radio to soft X-rays ($E \lesssim 10 \kev$). We include three sets of SED with different range of Eddington ratio $\lambda\equiv L_{\rm bol}/L_{\rm Edd}$ from \citet{Ho08}, i.e. $\lambda < 10^{-3}$, $10^{-3} <\lambda < 10^{-1}$, and $10^{-1} <\lambda < 1$. For comparison, the composite SED averaged over Type 1 and Type 2 AGNs compiled by \citet{Sazonov04} is also shown here by the black solid curve.

We caution that the origin of the nuclear infrared (IR) emission is rather complicated, i.e. it may come from the dusty torus, the circum-nuclear star formation, the central AGN (including the accretion disk, the jet, and sometimes the narrow line emission clouds), or their combination. Spatial resolution is thus of crucial importance to discriminate the contaminations, and extensive efforts have been made through infrared interferometric techniques (e.g., \citealt{Gandhi09, Tristram09, Asmus11, Asmus14, GM15, GM17}). However, these contaminations are still difficult to constrain (e.g. \citealt{Asmus11, Asmus14}). The nuclear IR flux derived from arcsecond-scale resolution observations (e.g. typical resolution in mid-IR of {\it Spitzer} is $\sim 4''$) may be accurate within a factor of $\lesssim 2-8$ \citep{Asmus11, Asmus14, GM17}.

%Interestingly, \citet{Gandhi09}, from a high resolution (sub-arcsecond) mid-IR VLT/VISIR observations of AGNs and found a universal and tight correlation between mid-IR luminosity and X-ray luminosity, i.e. ${\rm Log\ } L_{\rm MIR}= (0.41\pm0.03)+(1.12\pm0.04)\ {\rm Log\ } L_{\rm X}$, which as argued can be applied to both bright and faint AGNs.

%However, nearly all the existing data in IR are of poor resolution, i.e. the highest spatial resolution of current facilities is achieved by VLTI/VISIR is sub-arcsecond. Note that one arcsecond equals 326 and 600 pc, respectively, at redshift 0.016 (a medium value in Amus et al. 2011) and redshift $0.03$, thus inadequate to separate different components possible exist. Expect bright AGNs with luminosities $L>10^{44}\ergs$, Amus et al. (2011, 2014) found that the from sub-arcsecond-resolution observations of less luminous AGNs are much lower (by a factor of $2 - 8$) than their corresponding arcsecond-resolution Spitzer IRS observations. We will discuss the infrared data later in Sec.\ \ref{results}.

The spectrum from the hard X-ray to soft $\gamma$-ray regime ($10 \kev\lesssim E \lesssim 2\ E_c$) is absent in these composite SED data. Therefore, we complete the SED of this energy range through Eq. \ref{eq:xsed} (normalised at $E=10 \kev$), based on our discussions on $\Gamma$ and $E_c$ in \S\ref{sec:ecutoff}. Due to the uncertainties of $\Gamma$ and $E_c$, we choose different values of $\Gamma$ and $E_c$. As shown in Fig.\ \ref{fig:LLAGN_SED}, solid curves have relatively hard spectra, with $\Gamma = 1.60$, while the dashed curves have relatively soft spectrum, with $\Gamma = 1.80$. Different colors indicate different $e$-folding cutoff energies, i.e. dark green and dark red are for $E_c = 400\ \kev$ and $E_c = 800\ \kev$, respectively. Moreover, since the jet in LLAGNs are likely be relatively strong, we assume its emission be significant at $E\ga 2\ E_c$. We take the blazar spectrum from \citet{Sazonov04} (normalization is assumed to be 10\% the flux from hot accretion flow at $2\ E_c$) to consider the $\gamma$-ray emission in LLAGNs. Such an artificial jet emission in $\gamma$-ray band will affect the resultant Compton temperature in a minor way, i.e., less than $(3-5)\%$.

%Besides, the SED differences between Type 1 LLAGNs and Type 2 LLAGNs might not significant as their bright AGN cousins (Quasars or bright AGNs) who are distinctive in the optical and UV bands.

\section{Compton temperature of LLAGNs}\label{sec:comp}

\subsection{Method and equations}

Consider the scatter between photons with energy $\epsilon$ ($\epsilon \equiv h\nu/m_e c^2$) and electrons with temperature $T_e$ ($\theta_e\equiv kT_e/m_e c^2$). The heating or cooling rate of electrons is described by the following exact formulae which is valid for any photon energy and electrons temperature \citep{Guilbert86},
\begin{eqnarray}
q_{\rm Comp} & = & n_e\ \sigma_T\ \int {\sigma(\epsilon,\theta_e)\over\sigma_T}~ {\epsilon-<\epsilon_1>\over \epsilon}~F_\epsilon~d\epsilon \nonumber\\
& \equiv & n_e\ \sigma_T\ K_{\rm comp},\label{eq:qcomp}
\end{eqnarray}
where $F_\epsilon$ is the radiation flux at energy $\epsilon$, $K_{\rm comp}$ the kernel of the Compton heating rate. The cross-section for Compton scattering process has the form \citep{Guilbert86},
\begin{equation}
\sigma(\epsilon,\theta_e)=\frac{\sigma_T}{2K_2(1/\theta_e)}
\int^{+\infty}_{-\infty} g_0(\epsilon e^{\phi})\ e^{2\phi}\ {\rm
exp} \left(\frac{-{\rm cosh}~\phi}{\theta_e}\right)d\phi.\label{eq:sig}
\end{equation}
The average photon energy after scattering is \citep{Guilbert86},
\begin{eqnarray}
<\epsilon_1>&=&\epsilon+\frac{\sigma_T}{2K_2(1/\theta_e)\ \sigma}\int^{+\infty}_
{-\infty} \left(\theta_e+{\rm sinh}~\phi-\epsilon \right)\ \times \nonumber \\
& &\hspace{2cm} G(\epsilon
e^{\phi})\ e^{2\phi}\ {\rm exp}\left(\frac{-{\rm cosh~\phi}}{\theta_e}
\right)d\phi. \label{eq:ep1}
\end{eqnarray}
Here $K_2(x)$ is the 2nd order modified Bessel function, $G(\epsilon)\equiv g_0(\epsilon)-g_1(\epsilon)$ and
\begin{equation}
g_n(y) = \frac{3}{8}\int^2_0\left( t(t-2)+1+ty+\frac{1}{1+ty}
\right)\frac{dt}{(1+ty)^{n+2}}.
\end{equation}

In the Thompson limit ($h\nu\ll m_e c^2$ and $k T_e\ll m_e c^2$), Equations (\ref{eq:sig})\&(\ref{eq:ep1}) take their usual simple forms,
\begin{equation}
\sigma(\epsilon,\theta_e)\approx \sigma_T, ~~~~<\epsilon_1> \approx \epsilon+\epsilon\ (4\theta_e-\epsilon).
\end{equation}
In this case, by defining a ``Compton temperature'' $\tc$, Equation (\ref{eq:qcomp}) can be simplified as,
\begin{eqnarray}
q_{\rm Comp} & = & n_e {4k\sigma_T\over m_e c^2}\ F\ (\tc - T_e) \nonumber\\
 & = & n^2\ {n_e\over n}\ {k\sigma_T\over \pi m_e c^2}\ {L_{\rm bol}\over n R^2}\ (\tc - T_e) \label{eq:qcompsimp}
\end{eqnarray}
where $F\equiv\int F_\epsilon d\epsilon \equiv L_{\rm bol}/4\pi R^2$ is the radiative flux at distance $R$. The Compton temperature $\tc$ is defined as,
\begin{equation}
{k \tc\over m_e c^2}\equiv {1\over 4} {\int \epsilon F_\epsilon d\epsilon\over \int F_\epsilon d\epsilon}, \label{eq:tcdefine}
\end{equation}
i.e. it is the energy-weighted average energy of incident photons. From Equation\ \ref{eq:qcompsimp}, Compton scattering plays a heating (cooling) role when $\tc>T_e$ ($\tc<T_e$). In other words, $\tc$ is the temperature the gas at which net energy exchange by Compton scattering between photons and electrons vanishes. The electrons temperature of the ISM is usually $\la 10^7\ {\rm K}$, while generally $\tc$ is larger than $10^7\ {\rm K}$, so Compton scattering usually plays a heating role in the radiative feedback.

In the general case, the photon energy from AGNs can be comparable to or even larger than $m_ec^2$ and/or electrons can be relativistic; there is no exact definition of Compton temperature due to the strong coupling between electrons and photons. In this case, for the convenience of the calculation of Compton heating, we combine the exact Compton heating rate (Equation (\ref{eq:qcomp})) with its simplified version (Equation (\ref{eq:qcompsimp})) and define an ``effective'' Compton temperature as (see also \citealt{Sazonov04}),
\begin{eqnarray}
\tc & = & T_e + {m_e c^2\over k}\ {K_{\rm comp}\over \int b(\epsilon) F_\epsilon d\epsilon}\nonumber\\
 & \approx & T_e + {m_e c^2\over k}\ {K_{\rm comp}\over \int^{10 \kev}_0 F_\epsilon d\epsilon},\label{eq:tc}
\end{eqnarray}
where the correction factor $b(\epsilon)$ is unity for photon energy below $10\ \kev$ and decreases significantly above $10\ \kev$. With the evaluation of $\tc$, the Compton heating rate can be calculated easily by Equation (\ref{eq:qcompsimp}).

%r@{.}l fix the location of "."
\begin{table*}
\begin{center}
\vspace{0.2 cm}
\centerline{Table 2 -- Compton temperature of LLAGNs}\label{tab2}
\vspace{0.1 cm}	
\begin{tabular}{c | c c | c c c | c c c}	
\hline\hline
$\lambda$ & $\Gamma$ & $E_c$ & \multicolumn{6}{c}{$T_C/10^7\ {\rm K}$}\\
\cline{4-9}
($\equiv L_{\rm bol}/L_{\rm Edd}$) & & ($\kev$) & \multicolumn{3}{c}{normal case} & \multicolumn{3}{c}{reduced-IR case$^b$}\\
%\cline{4-9}
& & & ($T_e =10^4$ K) & ($T_e =10^5$ K) & ($T_e =10^6$ K) & ($T_e =10^4$ K) & ($T_e =10^5$ K) & ($T_e =10^6$ K)\\
\hline
\multirow{15}{*}{$\sim10^{-2}$}
& \multirow{5}{*}{1.50} & 300 & 4.66 & 4.66 & 4.65 & 9.40 & 9.40 & 9.38 \\
& & 400 & 5.64 & 5.64 & 5.62 & 11.4 & 11.4 & 11.4 \\
& & 500 & 6.52 & 6.52 & 6.50 & 13.2 & 13.2 & 13.2 \\
& & 600 & 7.29 & 7.29 & 7.27 & 14.7 & 14.7 & 14.7 \\
& & 700 & 8.00 & 7.99 & 7.98 & 16.1 & 16.1 & 16.1 \\
%\hline
\cline{2-9}
&  \multirow{5}{*}{{\bf 1.59}$^a$} & 300 & 3.44 & 3.44 & 3.43 & 6.95 & 6.95 & 6.93 \\
& & 400 & 4.10 & 4.10 & 4.09 & 8.28 & 8.28 & 8.26 \\
& & 500 & 4.68 & 4.68 & 4.67 & 9.45 & 9.45 & 9.43 \\
& & 600 & 5.19 & 5.19 & 5.18 & 10.5 & 10.5 & 10.5 \\
& & 700 & 5.64 & 5.64 & 5.63 & 11.4 & 11.4 & 11.4 \\
%\hline
\cline{2-9}
& \multirow{5}{*}{1.70} & 300 & 2.41 & 2.41 & 2.40 & 4.87 & 4.87 & 4.85 \\
& & 400 & 2.82 & 2.82 & 2.82 & 5.70 & 5.70 & 5.68 \\
& & 500 & 3.17 & 3.17 & 3.16 & 6.41 & 6.41 & 6.40 \\
& & 600 & 3.48 & 3.48 & 3.47 & 7.03 & 7.03 & 7.02 \\
& & 700 & 3.75 & 3.75 & 3.74 & 7.58 & 7.58 & 7.57 \\
\hline
\multirow{15}{*}{$<10^{-3}$}
& \multirow{5}{*}{1.60} & 400 & 7.36 & 7.36 & 7.34 & 14.1 & 14.1 & 14.1 \\
& & 500 & 8.37 & 8.37 & 8.35 & 16.1 & 16.1 & 16.0 \\
& & 600 & 9.27 & 9.27 & 9.25 & 17.8 & 17.8 & 17.8 \\
& & 700 & 10.1 & 10.1 & 10.1 & 19.4 & 19.4 & 19.3 \\
& & 800 & 10.8 & 10.8 & 10.8 & 20.8 & 20.8 & 20.7 \\
%\hline
\cline{2-9}
&  \multirow{5}{*}{{\bf 1.69}$^a$} & 400 & 5.65 & 5.65 & 5.63 & 10.9 & 10.8 & 10.8 \\
& & 500 & 6.37 & 6.36 & 6.35 & 12.2 & 12.2 & 12.2 \\
& & 600 & 6.99 & 6.99 & 6.97 & 13.4 & 13.4 & 13.4 \\
& & 700 & 7.54 & 7.54 & 7.52 & 14.5 & 14.5 & 14.4 \\
& & 800 & 8.04 & 8.04 & 8.02 & 15.4 & 15.4 & 15.4 \\
%\hline
\cline{2-9}
& \multirow{5}{*}{1.80} & 400 & 4.16 & 4.16 & 4.15 & 7.99 & 7.99 & 7.96 \\
& & 500 & 4.63 & 4.63 & 4.62 & 8.89 & 8.88 & 8.86 \\
& & 600 & 5.03 & 5.03 & 5.02 & 9.66 & 9.65 & 9.63 \\
& & 700 & 5.38 & 5.38 & 5.37 & 10.3 & 10.3 & 10.3 \\
& & 800 & 5.69 & 5.69 & 5.68 & 10.9 & 10.9 & 10.9 \\
\hline	
\end{tabular}
\end{center}
\small									
Notes: $^a$ the two photon indexes shown in bold, i.e. $\Gamma=1.59$ and $\Gamma=1.69$, are for cases with $L_{\rm bol}/L_{\rm Edd}= 10^{-2}$ and $10^{-3}$, respectively, where they are estimated from the $\Gamma-L_{\rm X}/L_{\rm Edd}$ relationship reported in \citet{Yang15} (see their Equation 5), with $L_{\rm X} = 1/16\ L_{\rm bol}$ \citep{Ho08}.\\
$^b$ the IR data, i.e. $\nu<3\times10^{14}$ Hz, in this case is reduced by a factor of $10$.
\vspace{0.1cm}
 \end{table*}

\subsection{Numerical results of $\tc$}

We now calculate the  Compton temperature of LLAGNs based on the SEDs shown in Figure\ \ref{fig:LLAGN_SED}. We consider two SEDs with different luminosity values, one being $10^{-3}<\lambda<10^{-1}$ and another $\lambda<10^{-3}$. They are simply represented as $\lambda\sim 10^{-2}$ and $\lambda<10^{-3}$ in Table\ 2. As discussed in \S \ref{sec:ecutoff}, the photon index and the cutoff energy of the X-ray spectrum of LLAGNs are currently poorly constrained. Therefore, for completeness, for each SED we have considered three values of $\Gamma$,  among which one is estimated from the $\Gamma-L_{\rm X}/L_{\rm Edd}$ relationship reported in \citet{Yang15} (see their Equation 5), with $L_{\rm X} = 1/16\ L_{\rm bol}$ \citep{Ho08}. For each $\Gamma$, we further consider five different values of $E_c$. Since the Compton heating rate also depends on the electron temperature of the ISM ($T_e$), we consider three different values of $T_e$, i.e., $T_e = 10^4, 10^5, 10^6$ K.

One point requires caution. For the calculation of $\tc$, the IR flux plays an important role since it largely determines the total flux, i.e., the denominator of Equation\ (\ref{eq:tc}). Observationally, however, as discussed in \S2, the origin of the infrared emission in the SED is not very clear. Due to the poor resolution of existing IR telescopes, the observed IR flux may only represent an upper limit \citep{Ho99, Asmus11, Asmus14, GM15, GM17}. Moreover, it is possible that the IR flux originates from a region far away from the black hole accretion flow. In the feedback study, we are often most interested in the region just beyond the outer boundary of the accretion flow, i.e., the Bondi radius $R_{\rm Bondi} = 2 G M_{\rm BH}/c_s^2 \approx 0.1\ {\rm kpc} \left(M_{\rm BH}/10^9\ \msun\right)\ \left(T_e/10^7\ {\rm K}\right)^{-1}$; this is because the properties of the gas in this region determines the black hole accretion rate, which subsequently determines the total output of the AGN (radiation, jet, and wind strength) and the growth of the black hole mass. So Compton heating in this ``inner'' region is most important. If the IR emission originates from far away from this region, their effect on the calculation of Compton temperature should not be included. Based on these considerations, we have also considered the ``reduced IR case'', in which the IR flux shown in Figure\ \ref{fig:LLAGN_SED} is artificially reduced by a factor of $10$ \citep{Asmus11, Asmus14, GM17}.\footnote{When the temperature of ISM is very low, e.g., the black hole is fueled by strong cooling flow, the Bondi radius can be as large as kpc scale or even larger \citep{Ciotti07}.  In this case, we should not consider the ``reduced-IR case'' if the IR emission region is within such a large Bondi radius. But note that the feedback in this case will be in radiative mode rather than kinetic mode.} Due to the reduction in both the bolometric luminosity and the Compton cooling from IR photons, this will make the Compton temperature higher by a factor of $\sim 2 - 3$.

%We note that the electrons in ISM and intergalactic medium are likely not that hot, i.e. $k T_e\ll m_e c^2$ or equivalently $T_e \ll 5\times 10^9$ K.

All these models and their Compton temperatures are listed in Table\ 2. From this table, the main results can be summarized as follows.
\begin{itemize}

\item The Compton temperature will be higher when the X-ray spectrum is harder or the cutoff energy is higher. For given $\Gamma$ and $T_e$, $\tc$ for cases with $E_c = 800\ \kev$ is a factor of $\sim 1.4$ higher than that for cases with $E_c = 400\ \kev$. The impact of $\Gamma$ is more evident, i.e. changing $\Gamma$ from $1.80$ to $1.60$ will result in a factor of $\sim 1.8$ increase in $\tc$. This is because a harder spectrum and a higher $E_c$ implies that there will be relatively more hard photons which can heat the electrons more efficiently than soft photons.
\item The value of $\tc$ is not sensitive to the temperature of ISM, $T_e$, as expected.
\item For reasonable parameter choices, the Compton temperature of LLAGNs with ``normal'' IR flux lies in the range $(5-9)\times 10^7\ {\rm K}$.
\item For the case of ``reduced-IR'' case, which is more favorable to us, the Compton temperature $\tc\approx (1-1.5)\times 10^8{\rm K}$.
\end{itemize}

Besides this statistical investigation of $\tc$ for LLAGNs, we have also calculated the Compton temperature of several individual LLAGNs. Such investigation is a little too ambiguous, since most of these sources lack the hard X-ray and soft $\gamma$-ray observations to constrain the value of $E_c$. So we use the theoretical SED results for such information, which is obtained by the accretion flow modeling to the sources (see \citealt{Yuan14} for details of the accretion model of LLAGNs). We select several representative sources from \citet{YYH09}. The results are shown in Table\ 3. Again, we have tried three different electron temperatures. The Compton temperature is systematically higher than the ``normal case'' shown in Table\ 2, but more consistent with the results of the ``reduced-IR'' case. The reason is that the IR flux calculated from theoretical model is significantly weaker than that shown in Figure\ \ref{fig:LLAGN_SED}, since we only consider the radiation from accretion flow and jet. In other words, the Compton temperature derived here is applicable to the regions close to the black hole, as we argue above.

%r@{.}l fix the location of "."
\begin{table}
\begin{center}
\vspace{0.2 cm}
\centerline{Table 3 -- Compton temperature of individual LLAGNs}\label{tab3}
\vspace{0.1 cm}	
\begin{tabular}{c | c |c c c}	
\hline\hline
Sources & $L_{\rm X}/L_{\rm Edd}$  & \multicolumn{3}{c}{$T_C/10^7\ {\rm K}$}\\
  &  & ($T_e =10^4$ K) & ($T_e =10^5$ K) & ($T_e =10^6$ K)\\
\hline
 NGC~4579  & $3.0\times10^{-4}$ & 25.9 & 25.9 & 25.9 \\
 NGC~6251  & $5.0\times10^{-5}$ & 9.85 & 9.85 & 9.83 \\
 NGC~4203  & $1.8\times10^{-5}$ & 16.1 & 16.1 & 16.1 \\
 NGC~~~315 & $1.5\times10^{-6}$ & 5.23 & 5.23 & 5.21 \\
 NGC~4296  & $1.3\times10^{-6}$ & 8.91 & 8.90 & 8.89 \\
 NGC~4594  & $1.2\times10^{-7}$ & 9.80 & 9.80 & 9.78 \\
% M~87  & $8.1\times10^{-8}$ & 0.673 & 0.673 & 0.672 \\
\hline																						
\end{tabular}
\end{center}
%\small									
Notes: observational data and theoretical modelling are taken from \citet{YYH09}. We caution that the hard X-rays and soft $\gamma$-rays in these sources are actually from SED modelling, as there are no direct observations at these energy bands.
%\vspace{0.1cm}
\end{table}

\section{The total radiative heating and cooling in LLAGNs}\label{sec:rad}

\begin{figure}
\centering
\includegraphics[width=8.5 cm]{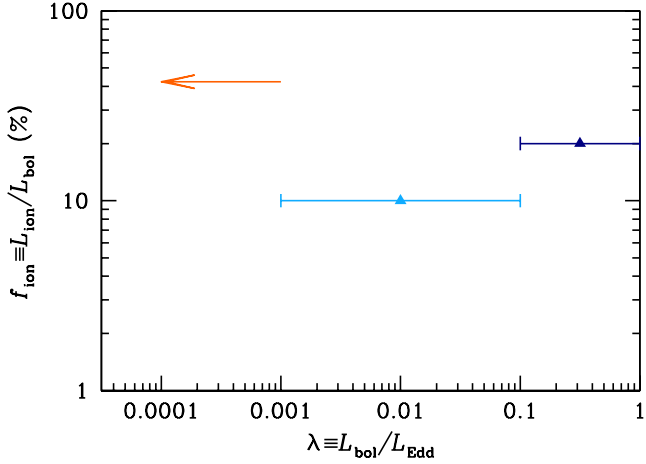}
\vspace{-0.1 cm}
\caption{The ionization luminosity factor, i.e. $f_{\rm ion} \equiv L_{\rm ion}/L_{\rm bol}$, versus the Eddington ratio, for the composite SED from \citet{Ho08}. The color of each curve is the same to that of Fig.\ \ref{fig:LLAGN_SED}.}\label{fig:fion}
\end{figure}

In addition to Compton scattering, there are several additional radiative processes that can heat or cool the gas. In this section we compare their relative magnitude. One process is the bremsstrahlung radiation ($q_{\rm br}$ and $S_{\rm br}$), which always plays a cooling role. The other processes relate to the atomic energy-level transition, i.e. the photoionzation/recombination and line emission ($q_{\rm ph,rec,l}$ and $S_{\rm ph,rec,l}$). The total radiative heating/cooling rate per unit volume can be expressed as (e.g., \citealt{Proga00,Sazonov05}),
\begin{eqnarray}
q_{\rm rad} & = & q_{\rm Comp} + q_{\rm ph,rec,l} - q_{\rm br}\label{eq:qr1}\\
& = & n n_e S_{\rm Comp} + n n_e S_{\rm ph,rec,l} - n^2 S_{\rm br} \nonumber\\
& = & n^2\ \left[{n_e\over n}(S_{\rm Comp} + S_{\rm ph,rec,l}) - S_{\rm br}\right].\label{eq:qr2}
\end{eqnarray}
Here $n=\rho/m_p$ and $n_e$ are the number density of hydrogen atomics and electrons, respectively. For hot gas (i.e. temperature $\ga 10^4$ K) with solar abundance, $n_e\approx n$. Consequently the radiative heating/cooling rate can be expressed as,
\begin{equation}
q_{\rm rad}\approx n^2 (S_{\rm ph,rec,l} + S_{\rm Comp}-S_{\rm br}).
\end{equation}

The value of $S_{\rm ph,rec,l}$ depends on both the metallicity (fixed to solar abundance in this work) and the ionization parameter $\xi$. The definition of $\xi$ is,
\begin{eqnarray}
\xi =  {L_{\rm ion} \over n R^2} & =& 27.3\ {\rm erg~s^{-1}~cm}\ \left({f_{\rm ion}\over0.2}\right)\ \left({\lambda\over10^{-2}}\right)\ \left({M_{\rm BH}\over 10^8\ \msun}\right)\  \nonumber\\
& & \hspace{1.8cm}\times\left({n\over 0.1\ {\rm cm}^{-3}}\right)^{-1}\ \left({R\over 1\ {\rm kpc}}\right)^{-2}.\label{eq:xi}
\end{eqnarray}
Here $R$ is the distance to the central black hole, $L_{\rm ion} = \int_{13.6\ {\rm eV}}^{13.6\ {\rm keV}} L_E\ dE$ is the ionization luminosity, and $f_{\rm ion} \equiv L_{\rm ion}/L_{\rm bol}$ is ionization luminosity factor. As shown in Figure\ \ref{fig:fion}, we derive the their ionization luminosity factors, i.e. $f_{\rm ion}\approx 0.42, 0.10, 0.20$, respectively, for the composite SEDs shown in Figure\ \ref{fig:LLAGN_SED} with $\lambda <10^{-3}, 10^{-3} - 10^{-1}$ and $10^{-1} - 1$. Detailed formulae of $S_{\rm ph,rec,l}$ is provided in Appendix (cf Equation\ \ref{eq:phrec}). It is a heating (cooling) term when $\xi$ is large (small).

\begin{figure}
\centering
\includegraphics[width=8.5 cm]{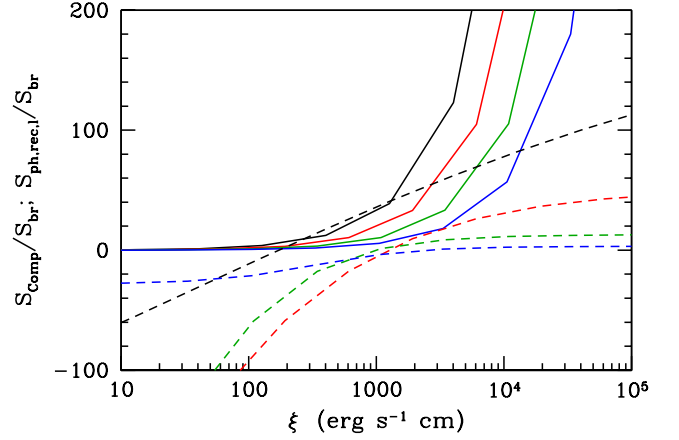}
\vspace{-0.1 cm}
\caption{The radiative heating/cooling rate in unit of the bremsstrahlung cooling rate, i.e. $S_{\rm Comp}/S_{\rm br}$ (solid curves) and $S_{\rm ph,rec,l}/S_{\rm br}$ (dashed curves), as a function of ionization parameter $\xi$. In this plot, we set $\tc=1\times10^8\ {\rm K}$. The color of each curve represents the temperature of electrons, i.e. $\log (T_e) = 4.5$ (black), $5.0$ (red), $5.5$ (green) and $6.0$ (blue).}\label{fig:hcratio}
\end{figure}

Figure\ \ref{fig:hcratio} provides the values of $S_{\rm Comp}/S_{\rm br}$ and $S_{\rm ph,rec,l}/S_{\rm br}$ as a function of $\xi$. In this figure, we set $\tc=1\times10^8\ {\rm K}$, and considered four different electron temperatures, i.e. $\log T_e = 4.5$ (black), $5.0$ (red), $5.5$ (green) and $6.0$ (blue). For this given electron temperature range, the Compton scattering plays a heating role. The total radiative process, on other hand, will play a cooling (heating) role for $\xi$ below (above) $\sim 10^{3}\  {\rm erg~s^{-1}~cm}$. When $\xi \ga 10^{4}\ {\rm erg~s^{-1}~cm}$, we have $S_{\rm Comp} > S_{\rm ph,rec,l}$.

\section{Summary}

This paper investigates the radiative heating in the kinetic mode of AGN feedback. This process is sometimes ignored in the AGN feedback study and people usually pay more attention to the kinetic feedback by the jet and wind. However, this may be over-simplified. Previous work in the case of the hard state of black hole X-ray binaries, in which we believe a hot accretion flow is operating just like in the kinetic mode, has shown that whenever the X-ray luminosity from the black hole  $L_{\rm X}\ga 4\times 10^{-5}L_{\rm Edd}$ (or roughly the bolometric luminosity $L_{\rm bol}\ga 6\times 10^{-4}L_{\rm Edd}$), the power of luminosity is larger than that of jet \citep{Fender03}. Depending on the accretion rate, the highest luminosity of the hot accretion flow in the kinetic mode can be as high as $(2-10)\% L_{\rm Edd}$. Moreover, compared to the AGN spectrum in the radiative mode, the spectral energy distribution of the LLAGNs in the kinetic mode is such that there are relatively more hard photons, which makes the radiative heating more effective. Based on these reasons, it is necessary to study systematically the radiative heating in the kinetic mode of AGN feedback.

This paper focus on Compton scattering. This process can in principle play a heating or cooling role, depending on the comparison between the photon energy and electron temperature of the gas. For this aim, we adopt the broad-band spectral energy distribution of LLAGNs with different luminosities. Based on this information, we have calculated the ``Compton temperature'' $\tc$, which characterizes the heating or cooling rate, and is the gas temperature at which the net energy exchange by Compton scattering between photons and electrons vanishes. Using this quantity, the Compton heating rate can be conveniently derived by Equation (\ref{eq:qcompsimp}) (see also Equations (\ref{eq:compa1})\&(\ref{eq:compa2}) in {\it Appendix}).

The results of $\tc$ are shown in Table\ 2, which gives $\tc\sim (5-15)\times 10^7{\rm K}$. This value is higher than the typical electron temperature of the gas in galaxies so it implies that in most cases, Compton scattering plays a heating role to the gas. This value is several times higher than the $\tc\approx 2\times 10^7{\rm K}$ value of luminous AGNs in the case of radiative mode of AGN feedback. The uncertainties in the $\tc$ of LLAGNs comes from two aspects. One is that the photon index and especially the cutoff energy of the hard X-ray spectrum of LLAGNs are poorly constrained. The other is that the exact value of the IR flux and more importantly, the distance at which we are evaluating the radiative heating is not well constrained. If we are considering the heating at a distance very far away from the black hole or accretion flow, we should adopt the lower value of $\tc\approx 5\times 10^7 {\rm K}$. But if we are interested in the heating not far away from the accretion flow, $\tc\approx 1.5\times 10^8{\rm K}$ should be adopted. We have also compared the Compton heating with the photoionization heating. We find that when the ionization parameter $\xi\ga 10^4$, Compton heating is larger than photoionization heating.

\acknowledgements
We thank Piotr Lubi\'nski for the useful discussions on measurements of the hard X-ray cutoff energy $E_c$, and Lei Hao and Y. Sophia Dai for useful discussions on IR observations of AGNs. We also appreciate Jerry Ostriker for helpful suggestions and comments. This work is supported in part by the National Program on Key Research and Development Project of China (Grant Nos. 2016YFA0400804, 2016YFA0400702 and 2016YFA0400704), the Youth Innovation Promotion Association of Chinese Academy of Sciences (CAS) (id. 2016243), the Natural Science Foundation of China (grants 11573051, 11633006 and 11661161012), the Natural Science Foundation of Shanghai (grant 17ZR1435800), and the Key Research Program of Frontier Sciences of CAS (No. QYZDJ-SSW-SYS008). This work has made extensive use of the NASA/IPAC Extragalactic Database (NED), which is operated by the Jet Propulsion Laboratory, California Institute of Technology, under contract with the National Aeronautics and Space Administration (NASA).

\begin{appendix}
\section{numerical formulae for the radiative heating/cooling terms}

Here we provide the numerical formulae for all the three radiative heating/cooling terms. The first is the Compton scattering. From Equation (\ref{eq:qcompsimp}) we have (see also \citealt{Proga00, Sazonov05}),
\begin{eqnarray}
S_{\rm Comp} &  = & 3.6\times 10^{-35}\ {L_{\rm bol}\over n R^2}\ (\tc - T_e)\ {\rm ergs\ s^{-1}\ cm^3} \label{eq:compa1}\\
 & = & 3.6\times 10^{-35}\ \left(\xi/f_{\rm ion}\right)\ (\tc - T_e)\ {\rm ergs\ s^{-1}\ cm^3}.\label{eq:compa2}
\end{eqnarray}

The second is the bremsstrahlung,
\begin{equation}
S_{\rm br} = 3.3\times10^{-27}\ T_e^{1/2}\ {\rm ergs\ s^{-1}\ cm^3}.\label{eq:br}
\end{equation}

The last includes photoionization heating, recombination cooling, and line cooling processes. We here take the detailed form from \citet{Sazonov05}, which is valid for $10^4\ {\rm K} \la T_e \la 3\times 10^7\ {\rm K}$,
\begin{equation}
S_{\rm ph,rec,l} = 10^{-23}\ {a+b(\xi/\xi_0)^c\over 1+(\xi/\xi_0)^c}\ {\rm ergs\ s^{-1}\ cm^3}.\label{eq:phrec}
\end{equation}
where
\begin{eqnarray}
\xi_0  & = & {1\over 1.5 T_e^{-0.5} + 1.5\times 10^{12}\ T_e^{-2.5}}  +{4\times10^{10}\over T_e^2}\ \left(1+{80\over e^{(T_e-10^4)/1.5\times10^3}}\right)~ {\rm erg~s^{-1}~cm}, \\
a & = & -{18\over e^{25\ (\log T_e - 4.35)^2}} - {80\over e^{5.5\ (\log T_e - 5.2)^2}} - {17\over e^{3.6\ (\log T_e - 6.5)^2}}, \\
b & = & 1.7\times 10^4\ T_e^{-0.7}, \\
c & = & 1.1 -{1.1\over e^{T_e/1.8\times10^5}} + {4.5\times10^{15}\over T_e^4}.
\end{eqnarray}

Another simplified version, which is not adopted in this work, is from \citet{Proga00} (note the difference in the definition of the Compton temperature), i.e.
\begin{eqnarray}
S_{\rm ph,rec,l} & = & 1.5\times 10^{-21}\xi^{1/4}T_e^{-1/2}\ (1-{T_e\over 4 \tc}) \nonumber\\
& & - \delta (10^{-24} + 1.7\times10^{-18}\ \xi^{-1}T_e^{-1/2}\ e^{-1.3\times10^5/T_e})\ {\rm ergs\ s^{-1}\ cm^3}.
\end{eqnarray}
Here parameter $\delta$ takes into account the effect of optical depth of lines; i.e. $\delta=1$ represents the optically thin line cooling case, and $\delta<1$ represents the case in which the line cooling is reduced when the lines becomes optically thick.

\end{appendix}

{}


\begin{thebibliography}{}

\bibitem[\protect\citeauthoryear{Asmus et al.}{2011}]{Asmus11} Asmus, D.,
    Gandhi, P., Smette, A., H\"{o}nig, S. F., Duschl, W. J., 2011, A\&A, 536, 36

\bibitem[\protect\citeauthoryear{Asmus et al.}{2014}]{Asmus14} Asmus, D.,
    H\"{o}nig, S. F., Gandhi, P., Smette, A., Duschl, W. J., 2014, \mnras, 439, 1648

\bibitem[\protect\citeauthoryear{Beckmann et al.}{2011}]{Beckmann11}
    Beckmann, V., Jean, P., Lubi\'nski, P., Soldi, S., Terrier, R., 2011, A\&A, 531, 70

\bibitem[\protect\citeauthoryear{Beckmann et al.}{2009}]{Beckmann09}
    Beckmann, V., Shrader, C.~R., Gehrels, N., et al.\ 2005, \apj, 634, 939

\bibitem[\protect\citeauthoryear{Beckmann et al.}{2005}]{Beckmann05}
    Beckmann, V., Soldi, S., Ricci, C. et al., 2009, A\&A, 505, 417

\bibitem[\protect\citeauthoryear{Bell et al.}{2011}]{Bell11} Bell, M.~E., Tzioumis, T., Uttley, P., et al., 2011, \mnras, 411, 402

\bibitem[\protect\citeauthoryear{Belloni}{2010}]{Belloni10} Belloni T. M., 2010, in Belloni T., ed., The Jet Paradigm ¨C From Microquasars to Quasars. Lecture Notes in Physics Vol. 794, Springer-Verlag, Berlin, p. 53

\bibitem[\protect\citeauthoryear{Blandford \& Begelman}{1999}]{BB99} Blandford, R. D., \& Begelman, M. C., 1999, \mnras, 303, L1

\bibitem[\protect\citeauthoryear{Blandford \& Znajek}{1977}]{Blandford77} 	Blandford, R. D., Znajek, R. L., 1977, \mnras, 179, 433

\bibitem[\protect\citeauthoryear{Bianchi et al.}{2004}]{Bianchi04} Bianchi S.,
    Matt G., Balestra I., Guainazzi M., Perola G.~C., 2004, A\&A, 422, 65

\bibitem[\protect\citeauthoryear{Brenneman et al.}{2014}]{Brenneman14}
    Brenneman L.~W., et al., 2014, \apj, 788, 61

\bibitem[\protect\citeauthoryear{Burke et al.}{2014}]{Burke14} Burke M.~J.,
    Jourdain E., Roques J.~P., Evans D.~A.,2014, \apj, 787, 50

\bibitem[\protect\citeauthoryear{Burlon et al.}{2011}]{Burlon11} Burlon, D.,
    Ajello, M., Greiner, J., Comastri, A., Merloni, A., Gehrels, N., 2011, \apj, 728, 58

\bibitem[\protect\citeauthoryear{Cheung et al.}{2016}]{Cheung16} Cheung, E.,
    Bundy, K., Cappellari, M., Peirani, S. et al. 2016, Natur., 533, 504

\bibitem[\protect\citeauthoryear{Choi et al.}{2012}]{Choi12} Choi, E., Ostriker, J. P., Naab, T., Johansson, P. H., 2012, \apj, 754, 125

\bibitem[\protect\citeauthoryear{Ciotti \& Ostriker}{2001}]{Ciotti01} Ciotti,
    L., \& Ostriker, J. P., 2001, \apj, 551, 131

\bibitem[\protect\citeauthoryear{Ciotti \& Ostriker}{2007}]{Ciotti07} Ciotti,     L., \& Ostriker, J. P., 2007, \apj, 665, 1038

%\bibitem[\protect\citeauthoryear{Ciotti \& Ostriker}{2012}]{Ciotti12} Ciotti, L., \& Ostriker, J. P. 2012, in Hot Interstellar Matter in Elliptical Galaxies, ed. D.-W. Kim \& S. Pellegrini, Springer-Verlag, Berlin, Astrophysics and Space Science Library, Vol. 378, p. 83

\bibitem[\protect\citeauthoryear{Ciotti, Ostriker \& Proga}{2010}]{Ciotti10} Ciotti, L., Ostriker, J. P., \& Proga, D. 2010, \apj, 717, 708

\bibitem[\protect\citeauthoryear{Crenshaw \& Kraemer}{2012}]{Crenshaw12}
    Crenshaw, D. M., \& Kraemer, S. B. 2012, \apj, 753, 75

\bibitem[\protect\citeauthoryear{Crenshaw, Kraemer \&
    George}{2003}]{Crenshaw03} Crenshaw, D. M., Kraemer, S. B., \& George, I.
    M. 2003, ARA\&A, 41, 117

\bibitem[\protect\citeauthoryear{Connolly et al.}{2016}]{Connolly16} Connolly, S., McHardy, I., Skipper, C., Emmanoulopoulos, D., 2016, \mnras, 459, 3963

\bibitem[\protect\citeauthoryear{Dadina}{2008}]{Dadina08} Dadina, M., 2008,
    A\&A, 485, 417

\bibitem[\protect\citeauthoryear{De Rosa et al.}{2007}]{DR07} De Rosa, A., Piro,
    L., Perola, G.~C., et al.\ 2007, A\&A, 463, 903

\bibitem[\protect\citeauthoryear{Denney et al.}{2006}]{Denney06} Denney, K.~D. Bentz, M~C., Peterson, B.~M., et al.\ 2006, \apj, 653, 152

\bibitem[\protect\citeauthoryear{Done et al.}{2007}]{Done07} Done, C., Gierli{\'n}ski, M., \& Kubota, A.\ 2007, \aapr, 15, 1

\bibitem[\protect\citeauthoryear{Eisenreich et al.}{2017}]{Eisenreich17} Eisenreich, M., Naab, T., Choi, E., Ostriker, J.P., Emsellem, 2017, MNRAS, 468, 751

\bibitem[\protect\citeauthoryear{Emmanoulopoulos et al.}{2012}]{Emm12}
    Emmanoulopoulos, D., Papadakis, I.~E., M$^c$Hardy, I.~M., Ar\'{e}valo, P.,
    Calvelo, D.~E., Uttley, P., 2012, \mnras, 424, 1327

\bibitem[\protect\citeauthoryear{Eracleous et al.}{2010}]{Eracleous10} Eracleous, M., Hwang, J.~A., Flohic, H.~M.~L., 2010, \apjs, 187, 135

\bibitem[\protect\citeauthoryear{Fabian}{2012}]{Fabian12} Fabian, A.~C.\ 2012,
    ARA\&A, 50, 455

\bibitem[\protect\citeauthoryear{Fender, Gallo \& Jonker}{2003}]{Fender03}
    Fender, R.P., Gallo, E., \& Jonker, P.G. 2003, MNRAS, 343, L99

\bibitem[\protect\citeauthoryear{F{\"u}rst et al.}{2016}]{Furst16} F{\"u}rst, F., M{\"u}ller, C., Madsen, K.~K., et al.\ 2016, \apj, 819, 150

\bibitem[\protect\citeauthoryear{Gebhardt et al.}{2000}]{Gebhardt00} Gebhardt, K., Bender, R., Bower, G., et al.\ 2000, \apjl, 539, L13

\bibitem[\protect\citeauthoryear{Gan et al.}{2017}]{Gan17} Gan, Z., Li, H., Li, S., Yuan, F., 2017, \apj, 839, 14

\bibitem[\protect\citeauthoryear{Gan et al.}{2014}]{Gan14} Gan, Z., Yuan, F.,
    Ostriker, J.~P., Ciotti, L., Novak, G.~S., 2014, \apj, 789, 150

\bibitem[\protect\citeauthoryear{Gandhi et al.}{2009}]{Gandhi09} Gandhi, P., Horst, H., Smette, A., et al.\ 2009, A\&A, 502, 457

\bibitem[\protect\citeauthoryear{Gondek et al.}{1996}]{Gondek96} Gondek, D.,
    Zdziarski, A.~A., Johnson, W.~N., et al.\ 1996, \mnras, 282, 646

\bibitem[\protect\citeauthoryear{Gonz\'{a}lez-Mart\'{i}n et al.}{2017}]{GM17}
    Gonz{\'a}lez-Mart{\'{\i}}n, O., Masegosa, J., Hern{\'a}n-Caballero, A., et al.\ 2017, \apj\ (in press), arXiv:1704.06739

\bibitem[\protect\citeauthoryear{Gonz\'{a}lez-Mart\'{i}n et al.}{2015}]{GM15}
    Gonz{\'a}lez-Mart{\'{\i}}n, O., Masegosa, J., M{\'a}rquez, I., et al.\ 2015, A\&A, 578, 74

\bibitem[\protect\citeauthoryear{Guilbert}{1986}]{Guilbert86} Guilbert, P.~W.,
    1986, \mnras, 218, 171

\bibitem[\protect\citeauthoryear{Guo}{2016}]{Guo16} Guo, F. 2016, \apj, 826, 17

\bibitem[\protect\citeauthoryear{Guo \& Mathews}{2011}]{Guo11} Guo, F., Mathews, W. G., 2011, \apj, 728, 121

\bibitem[\protect\citeauthoryear{Haehnelt \& Rees}{1993}]{HR93} Haehnelt, M.~G., \& Rees, M.~J., 1993, \mnras, 263, 168

\bibitem[\protect\citeauthoryear{Harris et al.}{2010}]{Harris10} Harris, G.~L.~H., Rejkuba, M., \& Harris, W.~E.\ 2010, PASA, 27, 457

\bibitem[\protect\citeauthoryear{Heckman \& Best}{2014}]{HB14} Heckman, T.~M., Best, P.~N., 2014, ARA\&A, 52, 589

%\bibitem[\protect\citeauthoryear{Hickox et al.}{2009}]{Hickox09} Hickox, R. C., Jones, C., Forman, W. R., et al.\ 2009, \apj, 696, 891

\bibitem[\protect\citeauthoryear{Ho}{1999}]{Ho99} Ho, L.~C., 1999, \apj, 516,
    672

\bibitem[\protect\citeauthoryear{Ho}{2008}]{Ho08} Ho, L.~C., 2008, ARA\&A, 46, 475

\bibitem[\protect\citeauthoryear{Homan et al.}{2016}]{Homan16} Homan, J.,
    Neilsen, J., Allen, J. L. et al. 2016, \apj, 830, L5

\bibitem[\protect\citeauthoryear{Honig et al.}{2014}]{Honig14} Honig, S.~F.,
    Watson, D., Kishimoto, M., Hjorth, J., 2014, Natur., 515, 528

\bibitem[\protect\citeauthoryear{Kauffmann \& Haehnelt}{2000}]{Kauffmann00} Kauffmann, G., Haehnelt, M., 2000, \mnras, 311, 576

\bibitem[\protect\citeauthoryear{Kormendy \& Ho}{2013}]{KH13} Kormendy, J.,
    Ho, L. C., 2013, \araa, 51, 511

\bibitem[\protect\citeauthoryear{Li, Ostriker \& Sunyaev}{2013}]{Li13} Li, J., Ostriker, J. P., Sunyaev, R., 2013, \apj, 767, 105

\bibitem[\protect\citeauthoryear{Lobban et al.}{2010}]{Lobban10} Lobban, A.~P., Reeves, J.~N., Porquest, D., Braito, V., Markowitz, A., Miller, L., Turner, T.~J., 2010, \mnras, 408, 551

\bibitem[\protect\citeauthoryear{Lubi\'nski et al.}{2016}]{Lubinski16} Lubi\'nski,
    P., Beckmann, V., Gibaud, L., et al.\ 2016, \mnras, 458, 2454

\bibitem[\protect\citeauthoryear{Lubi\'nski et al.}{2010}]{Lubinski10} Lubi\'nski,
    P., Zdziarski, A.~A., Walter, R., et al.\ 2010, \mnras, 408, 1851

\bibitem[\protect\citeauthoryear{Magorrian et al.}{1998}]{Magorrian98} Magorrian, J., Tremaine, S., Richstone, D., et al.\ 1998, \aj, 115, 2285

\bibitem[\protect\citeauthoryear{Maisack et al.}{1993}]{Maisack93} Maisack M.,
    et al., 1993, \apj, 407, L61

\bibitem[\protect\citeauthoryear{Malizia et al.}{2003}]{Malizia03}  Malizia, A.,
    Bassani, L., Stephen, J.~B., Coco, G.~D., Fiore, F., Dean, A.~J., 2003, \apj, 589, L17

\bibitem[\protect\citeauthoryear{Malizia et al.}{2008}]{Malizia08} Malizia, A., Bassani, L., Bird, A.~J., et al.\ 2008, \mnras, 389, 1360

\bibitem[\protect\citeauthoryear{Malizia et al.}{2012}]{Malizia12}Malizia A.,
    Bassani L., Bazzano A., et al.\ 2012, \mnras, 426, 1750

\bibitem[\protect\citeauthoryear{Malizia et al.}{2014}]{Malizia14} Malizia, A.,
    Molina, M., Bassani, L., et al.\ 2014, \apj, 782, L25

\bibitem[\protect\citeauthoryear{Marinucci et al.}{2014}]{Marinucci14} Marinucci, A., Matt, G., Kara, E., et al.\ 2014, \mnras, 440, 2347

\bibitem[\protect\citeauthoryear{Martini \& Weinberg}{2001}]{MW01} Martini, P., Weinberg, D.~H., 2001, \apj, 547, 12

\bibitem[\protect\citeauthoryear{Matt et al.}{2015}]{Matt15} Matt, G., Balokovi{\'c}, M., Marinucci, A., et al.\ 2015, \mnras, 447, 3029

\bibitem[\protect\citeauthoryear{McClintock \& Remillard}{2006}]{MR06} McClintock, J. E., Remillard, R. A., 2006, in LewinW. H. G., van der Klis M., eds, Compact Stellar X-ray Sources.Cambridge Univ. Press, Cambridge, p. 157

\bibitem[\protect\citeauthoryear{Miyakawa et al.}{2008}]{Miyakawa08} Miyakawa, T., Yamaoka, K., Homan, J., et al.\ 2008, PASJ, 60., 637

\bibitem[\protect\citeauthoryear{Molina et al.}{2009}]{Molina09} Molina, M.,
    Bassani, L., Malizia, A., et al.\ 2009, \mnras, 399, 1293

\bibitem[\protect\citeauthoryear{Molina et al.}{2013}]{Molina13} Molina, M.,
    Bassani, L., Malizia, A., et al.\ 2013, \mnras, 433, 1687

\bibitem[\protect\citeauthoryear{Murray et al.}{1995}]{Murray95} Murray, N.,
    Chiang, J., Grossman, S. A., Voit, G. M., 1995, \apj, 451, 498

\bibitem[\protect\citeauthoryear{Narayan \& Yi}{1994}]{NY94} Narayan, R., Yi, I., 1994, \apjl, 428, 13

%\bibitem[\protect\citeauthoryear{Narayan \& Yi}{1995}]{NY95} Narayan, R., \& Yi, I.\ 1995, \apj, 444, 231

\bibitem[\protect\citeauthoryear{Narayan et al.}{2012}]{Narayan12} Narayan, R., Sadowski, A., Penna, R.F., Kulkarni, A.K., 2012, \mnras, 426, 3241

\bibitem[\protect\citeauthoryear{Neill et al.}{2014}]{Neill14} Neill, J.~D., Seibert, M., Tully, R.~B., 2014, \apj, 792, 129

\bibitem[\protect\citeauthoryear{Neumayer}{2007}]{Neumayer07} Neumayer, N.,
    Cappellari, M., Reunanen, J., Rix H., van der Werf, P.~P., de Zeeuw, P.~T.,
    Davies, R.~I., 2007, \apj, 671, 1329

\bibitem[\protect\citeauthoryear{Onken et al.}{2014}]{Onken14}  Onken, C.~A., Valluri, M., Brown, J.~S., et al.\ 2014, \apj, 791, 37

\bibitem[\protect\citeauthoryear{Ostriker et al.}{2010}]{Ostriker10} Ostriker, J.P.
    Choi, E., Ciotti, L. et al. 2010, ApJ, 722, 642

\bibitem[\protect\citeauthoryear{Panessa et al.}{2008}]{Panessa08} Panessa, F., Bassani, L., de Rosa, A., et al.\ 2008, A\&A, 483, 151

\bibitem[\protect\citeauthoryear{Perola et al.}{2002}]{Perola02} Perola, G. C.,
    Matt, G., Cappi, M., et al.\ 2002, A\&A, 389, 802

\bibitem[\protect\citeauthoryear{Proga et al.}{2000}]{Proga00} Proga, D., Stone, J.~M., Kallman, T.~R., 2000, \apj, 543, 686

%\bibitem[\protect\citeauthoryear{Rejkuba}{2004}]{Rejkuba04} Rejkuba, M.\ 2004, \aap, 413, 903

\bibitem[\protect\citeauthoryear{Remillard \& McClintock}{2006}]{RM06}
    Remillard, R.~A., McClintock, J.~E., 2006, \araa, 44, 49

\bibitem[\protect\citeauthoryear{Sazonov, Ostriker \&
    Sunyaev}{2004}]{Sazonov04} Sazonov, S.~Y., Ostriker, J.~P., Sunyaev, R.~A.
    2004, \mnras, 347, 144

\bibitem[\protect\citeauthoryear{Sazonov et al.}{2005}]{Sazonov05} Sazonov,
    S.~Y., Ostriker, J.~P., Ciotti, L., Sunyaev, R.~A. 2005, \mnras, 358, 168

\bibitem[\protect\citeauthoryear{Shakura \& Sunyaev}{1973}]{SS73} Shakura, N.~I., Sunyaev, R.~A., 1973, A\&A, 24, 337

\bibitem[\protect\citeauthoryear{Stone, Pringle \& Begelman}{1999}]{Stone99} Stone, J. M., Pringle, J. E., \& Begelman, M. C. 1999, \mnras, 310, 1002

\bibitem[\protect\citeauthoryear{Soldi et al.}{2014}]{Soldi14} Soldi, S., Beckmann, V., Baumgartner, W.~H., et al.\ 2014, A\&A, 563, 57

\bibitem[\protect\citeauthoryear{Tombesi et al.}{2014}]{Tombesi14} Tombesi, F., Tazaki, F., Mushotzky, R.~F., et al.\ 2014, \mnras, 443, 2154

\bibitem[\protect\citeauthoryear{Tristram et al.}{2009}]{Tristram09} Tristram, K.
    R. W., Raban, D., Meisenheimer, K., et al.\ 2009, A\&A, 502, 67

\bibitem[\protect\citeauthoryear{Ursini et al.}{2015}]{Ursini15} Ursini, F., Marinucci, A., Matt, G., et al.\ 2015, \mnras, 452, 3266

\bibitem[\protect\citeauthoryear{Ursini et al.}{2016}]{Ursini16} Ursini, F., Petrucci, P.~-O., Matt, G., et al., 2016, \mnras, 463, 382

\bibitem[\protect\citeauthoryear{Vernaleo \& Reynolds}{2006}]{Vernaleo06} Vernaleo, J.~C., \& Reynolds, C.~S.\ 2006, \apj, 645, 83

\bibitem[\protect\citeauthoryear{Walsh et al.}{2013}]{Walsh13} Walsh, J.~L., Barth, A.~J., Ho, L.~C., Sarzi M., 2013, \apj, 770, 86

\bibitem[\protect\citeauthoryear{Weinberger et al.}{2017}]{Weinberger17}
    Weinberger, R., Springel, V., Herquist, L. et al. 2017, \mnras, 465, 3291

\bibitem[\protect\citeauthoryear{Wilson \& Yang}{2002}]{WY02} Wilson, A.~S.,
    Yang, Y., 2002, \apj, 568, 133

\bibitem[\protect\citeauthoryear{Winter et al.}{2009}]{Winter09} Winter, L.~M.,
    Mushotzky, R.~F., Reynolds, C.~S., Tueller, J., 2009, \apj, 690, 1322

\bibitem[\protect\citeauthoryear{Xie \& Yuan}{2012}]{Xie12} Xie, F.~G., Yuan,
    F., 2012, \mnras, 427, 1580

\bibitem[\protect\citeauthoryear{Yang et al.}{2015}]{Yang15} Yang, Q.~X., Xie,
    F.~G., Yuan, F., Zdziarsksi, A.~A., Ho, L.~C., Yu, Z., 2015, \mnras, 447, 1692

\bibitem[\protect\citeauthoryear{Yuan}{2016}]{Yuan16} Yuan, F., 2016, in
    Astrophysics of Black Holes, Astrophysics and Space Science Library, Vol.
    440, Springer-Verlag Berlin Heidelberg, p. 153

\bibitem[\protect\citeauthoryear{Yuan, Bu \& Wu}{2012}]{Yuan12} Yuan, F., Bu, D., Wu, M., 2012, \apj, 761, 130

\bibitem[\protect\citeauthoryear{Yuan et al.}{2015}]{Yuan15} Yuan, F., Gan, Z., Narayan, R., Sadowski, A., Bu, D., Bai, X., 2015, \apj, 804, 101

\bibitem[\protect\citeauthoryear{Yuan \& Narayan}{2014}]{Yuan14} Yuan, F.,
    Narayan, R., 2014, ARA\&A, 52, 529

\bibitem[\protect\citeauthoryear{Yuan, Yu \& Ho}{2009}]{YYH09} Yuan, F., Yu,
    Z., Ho, L.~C., 2009, \apj, 703, 1034

\bibitem[\protect\citeauthoryear{Zdziarski et al.}{1995}]{Zdziarski95} Zdziarski,
    A.~A., Johnson, W.~N., Done, C., Smith, D., McNaron-Brown, K., 1995, \apj, 438,
    L63

\bibitem[\protect\citeauthoryear{Zdziarski, Poutanen \& Johnson}{2000}]
    {Zdziarski00} Zdziarski, A.~A., Poutanen, J., Johnson, W.~N., 2000, \apj, 542,
    703

\end{thebibliography}
\end{document}